\documentclass[12pt]{iopart}

\usepackage[cp1251]{inputenc}
\usepackage{iopams}
\usepackage{graphicx,epsfig}
\begin{document}

\title[]{A variational principle, wave-particle duality, and the Schr\"{o}dinger equation}

\author{N. L. Chuprikov}

\address{Tomsk State Pedagogical University, 634041, Tomsk, Russia}
\ead{chnl@tspu.edu.ru} \vspace{10pt}

\begin{abstract}
A principle is proposed according to which the dynamics of a quantum particle in a one-dimensional configuration space
(OCS) is determined by a variational problem for two functionals: one is based on the mean value of the Hamilton operator,
while the second one is based on the mean value of the total energy of the particle, which is determined through the phase
of the wave function with help of the generalized Planck-Einstein relation. The first functional contains information
about the corpuscular properties of a quantum particle, and the second one comprises its wave properties. The true
dynamics is described by a wave function for which the variations of these two functionals are equal. This variational
principle, which can also be viewed as a mathematical formulation of wave-particle duality, leads to the Schr\"{o}dinger
equation.
\end{abstract}


\newcommand{\ppp}{\mbox{\hspace{5mm}}}
\newcommand{\ppa}{\mbox{\hspace{15mm}}}
\newcommand{\ppb}{\mbox{\hspace{20mm}}}
\newcommand{\ooo}{\mbox{\hspace{3mm}}}
\newcommand{\ooa}{\mbox{\hspace{1mm}}}

\section{Introduction: the Schr\"{o}dinger equation and the principle of least action} \label{int}

\hspace*{\parindent} As is known, the Schrodinger equation cannot be derived on the basis of the axioms of standard
quantum mechanics and, therefore, is considered by most researchers as its independent postulate. However, such a state of
affairs, when one of the main postulates of a physical theory is formulated in the form of a (albeit brilliantly guessed)
mathematical equation that does not have a generally accepted physical interpretation, is unacceptable for a physical
theory. In this regard, many attempts were made to formulate a physical principle that, on the one hand, would make it
possible to derive this equation and fill it with physical content, and, on the other hand, would not contradict the
already known principles of quantum mechanics.

Among the most well-known directions in solving this problem are the ``hydrodynamic formulation'' of Madelung-de
Broglie-Bohm, based on the concept of ``quantum fluid'' (here we mention only the recent work \cite{Mita}), as well as the
``quantum-field'' approach (see, for example, \cite{Zai,Der0,Der}), where the Schr\"{o}dinger equation is treated as a
classical field subject to quantization. In the ``quantum-field'' approach, this equation is derived from the principle of
least action, and it is important to understand to what extent this principle, which plays a fundamental role in classical
mechanics, respects the quantum mechanical principles.

As is known (see \cite{Zai,Der0}), in the quantum-field approach the Schr\"{o}dinger equation (for simplicity we restrict
ourselves to the one-dimensional case) appears as the Euler equation
\begin{eqnarray}\label{100}
\fl \frac{\partial {\cal{L}}}{\partial \psi}-\frac{\partial}{\partial x}\left(\frac{\partial {\cal{L}}}{\partial
\psi_x}\right)- \frac{\partial}{\partial t}\left(\frac{\partial {\cal{L}}}{\partial \psi_t}\right)=0
\end{eqnarray}
in the variational problem for the Lagrangian density
\begin{eqnarray}\label{101}
\fl {\cal{L}}=\frac{i\hbar}{2}\left(\psi^*\psi_t-\psi\psi_t^*\right)- \frac{\hbar^2}{2m}\psi_x^*\psi_x-V(x)\psi^*\psi;
\end{eqnarray}
here $\psi(x,t)$ is the wave function, and $\psi_x$ and $\psi_t$ are its $x$- and $t$-derivatives, respectively; $V(x)$ is
the potential of the external field in which the particle moves.

As we can see, the derivatives $\psi_x$ and $\psi_t$ enter this equation in the same way, and, as shown in \cite{Der}, in
this problem one can construct a consistent mathematical model of canonical quantization , where the time $t$ is equipped
with the corresponding operator . However, what is acceptable for canonical quantization is unacceptable for
nonrelativistic quantum mechanics, where $t$ is a parameter. Thus, it is impossible to introduce Hamilton's principle
(instead of postulating the Schr\"{o}dinger equation) into standard quantum mechanics without affecting its axioms.

In this article, we will show that it is possible to formulate a modified variational principle that not only leads to the
Schr\"{o}dinger equation (without violating other principles of quantum mechanics), but also allows us to deepen the
concept of wave-particle duality. As is known, this concept reflects one of the fundamental properties of a quantum
particle. But at present, this property is formulated as a general philosophical idea that allows mathematical formulation
-- in the form of the de Broglie relation and the Planck-Einstein relation - only in the case of single de Broglie waves.
In this article, we propose a mathematical formulation of wave-particle duality for an arbitrary (pure) quantum mechanical
state. It is this formulation that will form the basis of the modified variational principle, from which the
Schr\"{o}dingr equation will be derived.

\section{Generalized Planck-Einstein relation} \label{new}

Let the state of a particle moving in a one-dimensional configuration space (OCS) under an external field $V(x,t)$ be
described by the wave function
\begin{eqnarray}\label{2}
\fl \psi(x,t)=R(x,t)\ooa e^{i\phi(x,t)},
\end{eqnarray}
where $R(x,t)$ and $\phi(x,t)$ are the amplitude and phase of the wave function; $\int_{-\infty}^\infty R^2(x,t)dx=1$. It
is assumed that $\psi(x,t)$ belongs to the Schwartz space, the space of infinitely differentiable functions equal to zero
at infinity together with their derivatives.

In the Lagrangian formulation of classical mechanics, the Lagrange function for a particle in a one-dimensional space with
a generalized coordinate $q$ is defined as the difference between the particle's kinetic energy $K$ and its potential
energy $V$:
\begin{eqnarray}\label{1}
\fl L=K-V\equiv 2K-E,
\end{eqnarray}
where $E=K+V$ is the total energy of the particle. For a quantum particle, we define the Lagrange function $L_c(t)$ as the
difference {\it of the average values} of its kinetic and potential energies: $L_c(t)=\langle\psi|
\hat{K}|\psi\rangle-\langle\psi| V|\psi\rangle$, where $\hat{K}=\hat{p}^2/2m$; $\hat{p}=-i\hbar \frac{d}{dx}$ -- particle
momentum operator. Thus, the action $S_c$, which describes the quantum dynamics of a particle between times $t_1$ and
$t_2$, is
\begin{eqnarray}\label{121}
\fl S_c=\int_{t_1}^{t_2} L_c(t)dt\equiv K_c-V_c,\ppp K_c=\int_{t_1}^{t_2} \langle\psi| \hat{K}|\psi\rangle dt ,\ppp
V_c=\int_{t_1}^{t_2} \langle\psi| V|\psi\rangle dt.
\end{eqnarray}

In turn, the average values of $\langle\psi| \hat{K}|\psi\rangle$ and $\langle\psi| V|\psi\rangle$ are associated with the
kinetic energy density
\begin{eqnarray}\label{102}
\fl {\cal{K}}_c(x,t)=-\frac{h^2}{2m} \psi^*\frac{\partial^2\psi}{\partial x^2}=\frac{h^2}{2m} \left[
\frac{\partial\psi^*}{\partial x} \frac{\partial\psi}{\partial x}- \frac{\partial}{\partial x} \left(\psi^*
\frac{\partial\psi}{\partial x}\right)\right]=\nonumber\\
\fl =\frac{\hbar^2}{2m}\left[R^2\left(\frac{\partial \phi}{\partial x}\right)^2 +\left(\frac{\partial R}{\partial
x}\right)^2-\frac{\partial}{\partial x} \left(R\frac{\partial R}{\partial x}\right)\right]
\end{eqnarray}
and potential energy density
\begin{eqnarray}\label{102a}
\fl {\cal{V}}_c(x,t)=V(x,t)\psi^*(x,t)\psi(x,t)=V(x,t)R^2(x,t).
\end{eqnarray}
The sum of these two functions gives the density of the particle's total energy  ${\cal{H}}_c(x,t)$:
\begin{eqnarray}\label{103}
\fl {\cal{H}}_c={\cal{K}}_c+{\cal{V}}_c=-\frac{h^2}{2m} \psi^*\frac{\partial^2\psi}{\partial x^2}+V\psi^*\psi.
\end{eqnarray}

Now the terms $K_c$ and $V_c$ in the action $S_c$ (see (\ref{121})) can be written as
\begin{eqnarray*}\label{107}
\fl K_c=\int_{t_1}^{t_2}dt\int_{-\infty}^\infty dx\ooa {\cal{K}}_c(x,t);\ppp V_c=\int_{t_1}^{t_2}dt\int_{-\infty}^\infty
dx\ooa {\cal{V}}_c(x,t).
\end{eqnarray*}
And if we take into account the boundary conditions for the wave function and the fact that at times $t_1$ and $t_2$ the
(independent) variations of $\delta\psi$ and $\delta\psi^*$ are equal to zero, then it is easy to show that
\begin{eqnarray}\label{108}
\fl \delta K_c=\int_{t_1}^{t_2}dt\int_{-\infty}^\infty dx \left[-\frac{h^2}{2m}\frac{\partial^2\psi}{\partial
x^2}\right]\delta\psi^*+c.c.
\end{eqnarray}
\begin{eqnarray}\label{109}
\fl \delta V_c=\int_{t_1}^{t_2}dt\int_{-\infty}^\infty dx V\psi\delta\psi^*+c.c.
\end{eqnarray}
Consequently,
\begin{eqnarray}\label{110}
\fl \delta H_c=\delta K_c+\delta V_c=\int_{t_1}^{t_2}dt\int_{-\infty}^\infty dx
\left[-\frac{h^2}{2m}\frac{\partial^2\psi}{\partial x^2}+V\psi\right]\delta\psi^*+c.c.
\end{eqnarray}
\begin{eqnarray}\label{110a}
\fl \delta S_c=\delta K_c-\delta V_c=\int_{t_1}^{t_2}dt\int_{-\infty}^\infty dx
\left[-\frac{h^2}{2m}\frac{\partial^2\psi}{\partial x^2}-V\psi\right]\delta\psi^*+c.c.
\end{eqnarray}

And, at this point we come to the key point of the present approach. According to the principle of least action, in order
to determine the equation for the wave function, the variation $\delta S_c$ would have to be equated to zero. However,
based on Exp. (\ref{110a}), it is obviously impossible to obtain an equation describing the dynamics of the state of a
quantum particle, since this expression does not contain time derivatives. One more important argument against using this
principle of classical physics to derive the quantum mechanical equation is the fact that the variations (\ref{110a})
obtained on the basis of the operators of one-particle observables describe, in fact, only the corpuscular properties of a
quantum particle. To take into account its wave properties, it is necessary to turn to wave-particle duality, which
implies, in addition to $L_c$, another definition of the Lagrangian for a quantum particle.

Indeed, according to the Planck-Einstein and de Broglie relations, the wave $e^{i(kx-\omega t)}$ describes an ensemble of
(free) particles with energy $E=\hbar\omega$ and momentum $p=\hbar k$. For a wave function (\ref{2}) of a general form,
these relations can be written for each point of the OCS in the form of local connections between corpuscular and wave
characteristics:
\begin{eqnarray}\label{3}
\fl E(x,t)=\hbar\omega(x,t),\ooo p(x,t)=\hbar k(x,t);\\ \fl \omega(x,t)=-\frac{\partial\phi(x,t)}{\partial t},\ooo
k(x,t)=\frac{\partial\phi(x,t)}{\partial x}.\nonumber
\end{eqnarray}
It remains to clarify the question of whether the physical meaning of the quantities $E$ and $p$ is preserved in the
transition from the de Broglie wave to the wave function of the general form, that is, is it possible to treat the
function $E(x,t)R^2(x,t)$ as the particle's total energy density, and the function $p(x,t)R^2(x,t)$ as the particle's
momentum density.

The results obtained above allow us to unambiguously answer the second question. Indeed, if we take into account the
relation (\ref{3}) for the function $p(x,t)$ in Exp. (\ref{102}), then it becomes clear that the function $p(x,t)R^2(x
,t)$ determines only a part of the particle's kinetic energy and therefore cannot be interpreted as the momentum density
of a quantum particle. As regards the function $E(x,t)R^2(x,t)$, the results obtained above do not prohibit treating this
function as the density of the total energy of the particle. On this basis, we postulate that the wave function (\ref{2})
describes a one-particle quantum ensemble in which the  density of the total energy of particles at the OCS points is
given by the expression
\begin{eqnarray}\label{4}
\fl {\cal{H}}_w(x,t)=E(x,t)R^2(x,t)=-\hbar\frac{\partial\phi(x,t)}{\partial
t}R^2(x,t)\equiv\frac{i\hbar}{2}\left(\psi^*\frac{\partial\psi}{\partial t}-\psi\frac{\partial\psi^*}{\partial t}\right).
\end{eqnarray}

As a consequence, for a quantum particle, with making use of the right-hand expression in (\ref{1}) we can now define an
$S_w$ action that incorporates the wave properties of the particle: $S_w=2 K_c-H_w$, where
\begin{eqnarray}\label{111}
\fl H_w=\int_{t_1}^{t_2}dt\int_{-\infty}^\infty dx\ooa {\cal{H}}_w(x,t).
\end{eqnarray}
Since the (independent) variations of $\delta \psi$ and $\delta \psi^*$ are zero at times $t_1$ and $t_2$, it is easy to
show (integrating by parts) that
\begin{eqnarray}\label{112}
\fl \delta S_w=2 \delta K_c-\delta H_w,
\end{eqnarray}
where
\begin{eqnarray}\label{113}
\fl \delta H_w=\int_{t_1}^{t_2}dt\int_{-\infty}^\infty dx\ooa \delta{\cal{H}}_w(x,t)=i\hbar\int_{t_1}^{t_2}dt
\int_{-\infty}^\infty dx \left(\frac{\partial\psi}{\partial t} \delta \psi^* - \frac{\partial\psi^*}{\partial t} \delta
\psi\right).
\end{eqnarray}

Further, since the variations $\delta S_c$ and $\delta S_w$ (see Exps. (\ref{110}) and (\ref{112})) describe the same
particle, they must be equal. That is, the equality $2\delta K_c-\delta H_w=\delta K_c-\delta V_c$ must hold, from which
it follows that $\delta H_w=\delta K_c+\delta V_c\equiv \delta H_c$. In the last analysis, the essence of the proposed
variational approach is that the variation of the functional $H_c$, which describes the corpuscular properties of a
particle, must coincide with the variation of the functional $H_w$, which describes its wave properties:
\begin{eqnarray}\label{114}
\fl \delta H_c=\delta H_w.
\end{eqnarray}
Now we have here to take into account Exps. (\ref{110}) and (\ref{113}) for $\delta H_c$ and $\delta H_w$, and the fact
that the equality (\ref{114}) must hold for both independent variations of $\delta\psi$ and $\delta \psi^*$. Lastly,
equating the integrands containing the $\delta\psi^*$ variation, we obtain the Schr\"{o}dinger equation
\begin{eqnarray*}
\fl \frac{\partial\psi}{\partial t}=-\frac{h^2}{2m}\frac{\partial^2\psi}{\partial x^2}+V\psi.
\end{eqnarray*}

\section{Conclusion}

A variational principle has been developed, which, in fact, represents a mathematical formulation of wave-particle duality
for the wave function of a general form:
\begin{eqnarray*}
\fl \int_{t_1}^{t_2}dt\int_{-\infty}^\infty dx\ooa
\delta\left(\psi^*\hat{H}\psi\right)=\frac{i\hbar}{2}\int_{t_1}^{t_2}dt\int_{-\infty}^\infty dx\ooa
\delta\left(\psi^*\frac{\partial\psi}{\partial t}-\psi\frac{\partial\psi^*}{\partial t}\right).
\end{eqnarray*}
This principle does not conflict with the well-known postulates of standard quantum mechanics.

\end{document}